\documentclass{moriond}

\usepackage{amsmath,amssymb,mathtools}
\usepackage{graphicx}
\usepackage[utf8]{inputenc}
\usepackage{enumerate} 
\usepackage{slashed}
\usepackage{comment}
\usepackage{multirow}
\usepackage{acronym}
 \usepackage{bbold}

\acrodef{PDG}[PDG]{Particle Data Group}
\acrodef{OPE}[OPE]{Operator Product Expansion}
\acrodef{FCNC}[FCNC]{flavour-changing neutral current}
\acrodef{RHC}[RHC]{right-handed currents}
\acrodef{SM}[SM]{Standard Model}
\acrodef{NP}[BSM]{Beyond the \ac{SM}}
\acrodef{MFV}[MFV]{Minimal Flavour Violation}
\acrodef{SD}[SD]{short-distance}
\acrodef{LD}[LD]{long-distance}
\acrodef{DA}[DA]{distribution amplitude}

\newcommand{\CondQQ}[1]{{ \left\langle \bar{#1} #1 \right \rangle}}

\newcommand{\vev}[1]{\langle #1 \rangle}

\newcommand{\matel}[3]{\langle #1|#2|#3\rangle}
\newcommand{\al}{\alpha}
\newcommand{\be}{\beta}
\newcommand{\ga}{\gamma}
\newcommand{\de}{\delta}

\newcommand{\eps}{\epsilon}

\newcommand{\Tr}{{\textrm{Tr}}}

\newcommand{\amp}{ \mathcal A}
\newcommand{\eff}{\textrm{eff}}

\newcommand{\SU}{\text{SU}}

\newcommand{\U}{\text{U}}

\newcommand{\fone}{f_1}

\newcommand{\fonetwo}{f_1(\text{\small{1420}})}

\newcommand{\DD}{D} 

\newcommand{\HH}{H}

\newcommand{\RR}{\mathbb{R}}
\newcommand{\GG}{G}
\newcommand{\epsQCD}[1]{{\eps}^{#1}}

\newcommand{\Rea}{\textrm{Re}}
\newcommand{\Ima}{\textrm{Im}}

\def\al{\alpha}

\def\be{\begin{equation}}
\def\ee{\end{equation}}
\def\bea{\begin{eqnarray}}
\def\eea{\end{eqnarray}}

\newcommand{\Photo}{}

\begin{document}
\begin{flushright}
\begin{tabular}{l}
 CP3-Origins-2018-025 DNRF90
 \\
 \end{tabular}
\end{flushright}
\vskip1.5cm

\title{Right-handed Currents Searches and Parity Doubling\,\footnote{Proceedings of Moriond-QCD 2018.}}

\author{James Gratrex and Roman Zwicky}

\address{Higgs Centre for Theoretical Physics, School of Physics and Astronomy,\\
University of Edinburgh, Edinburgh EH9 3JZ, Scotland}

\maketitle

\abstracts{The extraction of right-handed currents, beyond the Standard Model, faces theoretical challenges 
from long-distance contributions. We show that these effects can be controlled by combining, for example, 
studies of $B \to V(1^-) \ga$ and $B \to A(1^+) \ga$ observables. The sum of the long-distance contributions can be extracted without compromise, and the individual pieces follow from a ratio predicted by theory. 
This leads to significant reduction in the uncertainty of long-distance contributions. The ideas extend to charm decays and the low $q^2$-region of $B \to V \ell \bar{\ell}$, and open the prospect of checking input affecting the angular $B \to K^* \mu \mu$-anomaly.} 

\section{Introduction}\label{subsec:prod}

The \ac{SM} is a highly successful, yet peculiar, theory. One of its peculiarities is that
the weak interactions are of the V-A type. It is intuitively clear that this leaves traces 
in  the polarisation (or the angular distribution) of weak decays. 
Such traces would be  perfect probes for \ac{RHC} searches were it not for non-perturbative  effects of QCD diluting the purity of the signal.

A particularly good setting to test the chirality of interactions is when there is a photon in the final
state, as the photon helicity is then in direct correspondence with the handedness of the interaction.
In particular,  in the limit of no (quark) masses, 
chirality and handedness are the same. For example, the QED interaction reads
$\bar q  \slashed{A} q  = \bar q_L  \slashed{A} q_L + \bar q_R  \slashed{A} q_R$. 
Thus, the reaction $\bar q_L + q_L \to \ga_L$ is an on-shell process where the two half-helicities of the 
quarks add up to match the $\pm 1$  helicity of the photon (termed left- and right-handed respectively). If the interactions 
were chiral, $H^{\textrm{int}} \sim \bar q_L  \slashed{A} q_L$, then the resulting photon polarisation would always be left-handed. 
Denoting the amplitude of left-   and right-handed photons by 
${\cal A}_{L,R}$ respectively, this reads ${\cal A}_R / {\cal A}_L = 0$.

Such transitions are not present in the \acp{FCNC} in the \ac{SM}.  
The next best possibility is to couple the photon to two quarks of opposite chirality in a
directly gauge-invariant way, at the expense of a quark mass term. This is realised by  the so-called electric dipole  operator (known as the  $O^{(')}_{7}  $-term in the effective Hamiltonian)
\begin{equation}
 H^{\textrm{eff}}  \supset  C^{(')}_7 O^{(')}_{7}  \,  \sim \, m_b(m_s)  {\bar s}_{L(R)}  \sigma_{\mu \nu}   F^{\mu \nu}  b_{R(L)}  \;, \quad %O'_{7}    \sim m_s  {\bar s}_R  \sigma_{\mu \nu}   F^{\mu \nu}  b_L  \;,
\end{equation}
where  $F^{\mu \nu} $ is the photon field strength tensor. The concrete appearance of $m_b(m_s)$ can be understood from a spurion analysis \cite{MFV}.  The dimension-six effective Hamiltonian is written as 
$H^{\textrm eff}_{b \to s \ga/\ell \bar \ell}  \sim C \bar s_L \Gamma b O_r + C' \bar s_R \Gamma b O_r $,
with  flavour-neutral $O_r$ and
$C'/C|_{SM} \ll 1$ similar to $C_7'/C_7|_{SM} = m_s/m_b$. This hierarchy, and therefore \ac{RHC} searches, is affected 
by the non-perturbative QCD matrix elements
\begin{equation}
\label{eq:15}
{\cal A}_{\mathbb{1}(\mathbb{5})}   = \matel{X_s \ga^*}{\bar s (\ga_5) \Gamma b O_r}{B} \;.
\end{equation}

 More concretely, one of the 
 simplest processes testing the helicity is the $B \to V \ga$ decay, where $V$ is a $J=1$ vector meson. 
 The amplitude of left- and right-handed polarised photon is then given by
 \begin{equation}
 \label{eq:ALR}
 {\cal A}_{R(L)}  \sim 
 C( {\cal A}_{\mathbb{1}} \mp {\cal A}_{\mathbb{5}} ) + 
 C'( {\cal A}_{\mathbb{1}} \pm {\cal A}_{\mathbb{5}} ) \;,
 \end{equation}
 making the link between chirality and photon polarisation explicit. 
  In that case, the algebraic relation 
 $\sigma^{\alpha \beta} \ga_5 = -  
 \frac{i}{2} \eps^{\al\beta\ga\de}\sigma_{\ga\de}$,
 does miracles,   ${\cal A}_{\mathbb{1}}/ {\cal A}_{\mathbb{5}}|_{O_7,O_7'} = T_1(0)/T_2(0) =1 $, 
 on the level of the dipole operator as it enforces the form factor relation $T_1(0)= T_2(0)$, see Ref.~\cite{BSZ2015} for example.
This then results in 
\begin{equation}
\frac{{\cal A}_R}{  {\cal A}_L} \Big|_{O_7,O^{'}_{7}}   =   \frac{m_s}{m_b} \equiv \hat{m}_s \; \;.
\end{equation}
\ac{NP} shifts to $O_7'$, of the form $\hat{m}_s \to \hat{m}_s + \Delta_{\textrm{RHC}}$, are what we refer 
to as \ac{RHC} in this context.
As hinted at, \ac{LD} contributions $\eps_{L,R}$ dilute the purity of the signal. Schematically, 
\begin{equation}
\label{eq:break1}
\frac{{\cal A}_R}{  {\cal A}_L}\Big|_{B \to V\ga} 
= \frac{ \eps_R + \hat{m}_s +  \Delta_{\textrm{RHC} }   }{1 +  \eps_L}  \simeq 
\eps_R +  \hat{m}_s +  \Delta_{\textrm{RHC} }    \;,
\end{equation}
where we have assumed that $\eps_L \ll 1$, which computations \cite{KSW1995,BZ06CP,BJZ2006}
and indirect evidence tend to support. 
Eq.~\eqref{eq:break1} makes it clear that in order to distinguish \ac{RHC} from \ac{LD} contributions one needs
to be able to predict $\eps_R$ (or  essentially  $ {\cal A}_{\mathbb{1}} - {\cal A}_{\mathbb{5}}$ up to ${\cal O}(C'/C)$). 

In this work, we advocate a novel approach invoking new observables \cite{prep_RHC}.
At the level of the standard dimension-six
 $H^{\textrm{eff}}$, $\eps_R$ arises through an $\bar s_L \Gamma b \,  \bar q \Gamma' q$-interaction, and is 
sensitive to the parity quantum number of the $V$-state. 
Hence, if for every vector state there were a partner of opposite parity then one could discern 
$\Delta_{\textrm{RHC}}$ from  $\eps_R$. This is the idea of our work, and we advocate 
to combine decay channels of nearly-degenerate parity partners.

In the chiral symmetry restoration limit, the following exact relation  
will be shown to hold:
\begin{equation}
\label{eq:Ampav}
\amp^{B \to V \ga}_\chi(C,C') =  \amp^{B \to A \ga}_\chi(-C,C') \;,
\end{equation}
with $\chi = L,R$, and $A$ an opposite-parity partner of the $V$ meson.
This is the solution to our problem, since we are concerned with mixing up $C'$ (i.e.  $\Delta_{\textrm{RHC}}$) with \ac{LD}-effects $\eps_R$ induced by  $C$-type operators cf. Eq.~\eqref{eq:ALR}. 
As we shall see, the problem 
is then shifted from estimating $ {\cal A}_{\mathbb{1}} - {\cal A}_{\mathbb{5}}$ to estimating
\begin{equation}
\label{eq:Rschema}
\RR_{A,V} \equiv  \frac{\textrm{Re} [\eps_R^{B \to A \ga} ]}{\textrm{Re} [\eps_R^{B \to V \ga} ]}  
= 1+ {\cal O}( m_q ,\vev{\bar qq}) \;.
\end{equation}
To some extent it is the operator state correspondence of quantum 
field theory, expressed by the LSZ formalism, that allows for this shift in perspective.
We would like to stress already at this point that 
the crucial practical question is not  the actual value of $\RR_{A,V}$, 
but rather its uncertainty (in the real world),
which of course indirectly benefits from the closeness to the symmetry limit.

\section{Relating axial and vector meson matrix elements in the chiral symmetry limit}

It is conceptually beneficial to consider the chiral restoration limit $\{  m_q , \vev{\bar q q} ,  \dots\} \to 0$, 
where the axial flavour symmetries  are restored:
$ \SU(N_F)_V \to \SU(N_F)_V \times \SU(N_F)_A \times \U(1)_A$.\footnote{To what extent the $U(1)_A$ is restored 
due to the axial anomaly is an interesting question, but is not relevant for our purposes. Finite-temperature
lattice computations above the chiral phase transition give evidence of $U(1)_A$-restoration \cite{DW}.} 
The relation we are to use is that  in the restoration limit  the quark propagator in the gluon background field,
  $S^{(q)}_\GG(w,z) = \matel{w}{(\slashed{D}  +i m_q )^{-1}} {z}$, obeys 
  \begin{equation}
\label{eq:ga5}
\ga_5 S^{(q)}_\GG(w,z) = - S^{(q)}_\GG(w,z) \ga_5 \;,
\end{equation}
for which the vanishing of the $ \SU(N_F)_A \times \U(1)_A$-violating condensates is a necessary condition, as can be understood from  
the Banks-Casher relation \cite{prep_RHC}.

The starting point is that any information of the matrix element
$\matel{V \ga^*}{ h_{\eff}}{B}$, where $\ga^*$ is a potentially  off-shell photon, can be extracted from the correlation function 
\begin{equation}
\label{eq:M}
{\cal M}^{[V]}_{(v,a)}%(x,y,0) 
\equiv  \matel{0}{\mathcal{T}\{ J_B(x) V^I_\mu(y) h_{\eff}(0) \}}{0 } \;,  \quad h_{\eff} = \bar q ( v + a \ga_5) \Gamma b \, O_r  \;,
\end{equation}
by analysing its dispersion relation, as  $J_B = \bar b \ga_5 q$ and 
$V^I_\mu \to \rho(a_1)^I_\mu = \bar q  \ga_\mu T^I (\ga_5) q$ are interpolating operators for
the $B$-meson and the vector (axial) mesons respectively.  In Eq.~\eqref{eq:M}, $\Gamma$ is a Dirac structure, while 
$O_r$ stands for the remaining part of the operator. For example, 
$O_r = \mathbb 1$ and $O_r = \bar c \Gamma' c,  \bar u \Gamma' u, \dots$  distinguish between 
\ac{SD} and \ac{LD} (e.g. four-quark) operators. Contracting the quark lines and focusing on the $\rho$ meson final state,
the matrix element assumes the form
\begin{equation}
\label{eq:p-integral}
{\cal M}^{[\rho^0]}_{(v,a)} \sim  \int D \mu_\GG \Tr[ (v+a \ga_5) S_G^{(b)}(0,x) \ga_5 S_\GG^{(d)}(x,y) \ga_\mu S_\GG^{(d)}(y,0)] \;,
\end{equation} 
where the path-integral measure is given by $D \mu_\GG = D \GG_\mu \det( \slashed{D} + i M_f) e^{i S(\GG)} $ ($ D_\mu = (\partial - ig G)_\mu$).  

\begin{figure}[t!]
\centering
\includegraphics[scale=0.6,clip=true,trim=20 520 0 70]{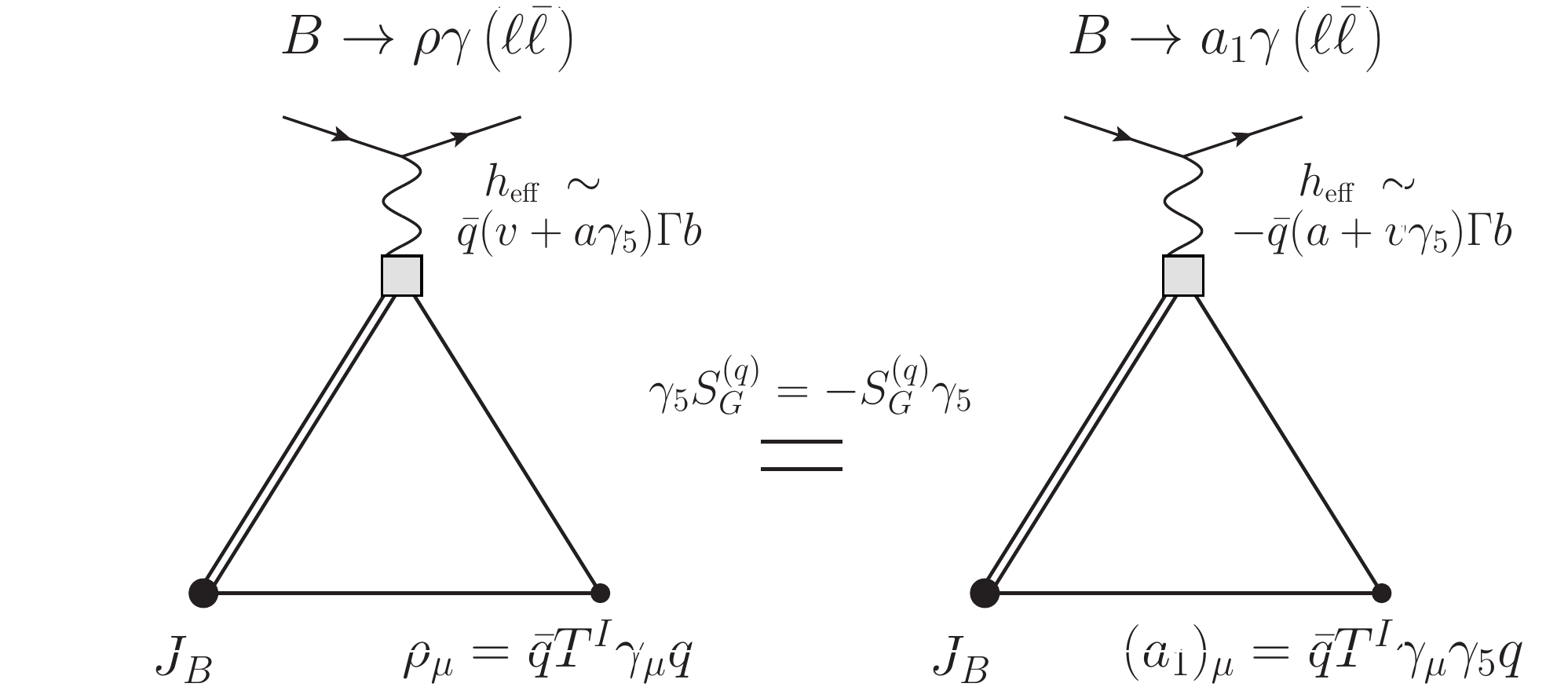}
\caption{\small A diagrammatic interpretation of the procedure outlined in the main text, using the relation Eq.~\eqref{eq:ga5}. which necessitates  the limits $\{ m_q,\CondQQ{q} ,\dots \} \to 0$. 
The argument only requires that
the weak vertex $h_{\textrm{eff}}$ be a local operator, 
and thus applies to both \ac{SD}  (form-factor) and \ac{LD} (charm-loop) contributions.  
Note that the trick applies equally well to annihilation diagrams where the photon is emitted from 
one of the lines inside the triangle graph in the figure. 
The schematic correlation functions on the left and right are exactly equal, from where the information on the matrix elements 
can be assessed. %Corrections to this exact equality, beyond the chiral symmetry limit, are discussed in  \cite{prep_RHC}.
}
\label{fig:PD_Diagram}
\end{figure}

Now comes the main trick. Substituting $ \ga_\mu \to  \ga_\mu (\ga_5)^2$ and using Eq.~\eqref{eq:ga5} leads to an expression, $ {\cal M}^{[a_1]}_{(a,v)}  = -  {\cal M}^{[\rho^0]}_{(v,a)}$, 
for which the $a_1$ meson matrix element is the same up to a sign with the variables $a$ and $v$ interchanged (Fig.~\ref{fig:PD_Diagram}). From this expression, Eq.~\eqref{eq:Ampav} follows, which 
is our main formal result.  In the last equation it is understood that $V$ and $A$ become degenerate 
in the chiral restoration limit \cite{prep_RHC}, and are referred to as parity doublers.\footnote{Parity doubling has a long history in particle physics 
\cite{Afonin:2007mj}, and has recently been investigated on the lattice \cite{DGL15,Rohrhofer:2017grg}, 
with the additional surprise of an emergent symmetry. For a table of relevant opposite parity states, we refer the reader to Tab.~1 in Ref.~\cite{prep_RHC}.}

\section{Phenomenological implications}

In effect, Eq.~\eqref{eq:Ampav}  means that the ratio
Eq.~\eqref{eq:break1}, amended to include an axial  meson, is then given by
\begin{equation}
\label{eq:break2}
\frac{{\cal A}_R}{  {\cal A}_L}\Big|_{B \to V(A)\ga} 
= \frac{\eps_R \pm(  \hat{m}_s +  \Delta_{\textrm{RHC} } ) }{(1 +  \eps_L)}  \simeq 
 \eps_R \pm(\hat{m}_s +  \Delta_{\textrm{RHC} })   \;,
\end{equation}
which is the equation from which our phenomenological results are derived. 
Beyond the symmetry limit, the only relevant change to Eq.~\eqref{eq:break2} is that the \ac{LD} contributions
$\eps_{R},\eps_L \to \eps_{V(A),R},\eps_{V(A),L} $  are dependent on the final state 
to a degree that needs to be estimated by analytical methods.

The crucial question for \ac{RHC} searches
is how the hierarchy and the sign change in Eq.~\eqref{eq:break2} 
can be exploited. Concerning the hierarchy, the rate itself is not promising, since
$\Gamma_{\textrm{tot}} \sim  |{\cal A}_L|^2  +   |{\cal A}_R|^2$ and the effect of the \ac{RHC}s might be
too small to be seen in experiment. A more promising route is to consider angular distributions, 
e.g. $B \to V \ell \bar{\ell}$, or time-dependent decay rates in $B \to V \ga$, as originally proposed 
by the authors of Ref.~\cite{AGS97}. 

The time-dependent rate of a neutral $B_{D}$-meson ($D=d,s$), under general and valid assumptions, reads \cite{MXZ2008}
\begin{equation}
\label{eq:bbar-t}
{\cal B}(\bar{B}_\DD[B_\DD] \to V \ga) = 
     B_0 e^{-\Gamma_\DD t}[{\textrm{ch}}(\frac{\Delta \Gamma_\DD}{2}t) -\HH { \textrm{sh}}(\frac{\Delta \Gamma_D}{2}t)
                \mp C\cos(\Delta m_{\DD} t) \pm S \sin(\Delta m_\DD t)] \;,
\end{equation}
where $\Delta \Gamma_\DD \equiv \Gamma^{(H)}_\DD- \Gamma^{(L)}_\DD $ is the width difference, and $\Delta m_\DD \equiv m^{(H)}_\DD- m^{(L)}_\DD $ the mass difference, of the heavy ($H$) and light ($L$)  mass eigenstates.  
$S$ and $C$ are related  to indirect and direct CP violation respectively.\footnote{  $H \equiv {\cal A}^{\Delta \Gamma}$ in the \ac{PDG} notation.} The quantities $S$ and $H$ are linear in $ \amp_R$, and  
 given in terms of the amplitudes by 
\begin{alignat}{2}
\label{eq:SHdef}
& S(H)  &\;=\;&   2   { \Ima(\Rea) }\left [\frac{q}{p}(\bar \amp_L \amp_L^* +
\bar \amp_R \amp_R^*) \right]  {\cal N}^{-1}   \; ,
 \end{alignat}
 with ${\cal N} =  |\amp_L|^2 + |\bar \amp_L|^2 
+  |\amp_R|^2 + |\bar \amp_R|^2$.
We choose to illustrate the approach by the mode $B_s \to \phi\ga$ and $B_s \to \fonetwo\ga$, with other
modes discussed in Ref.~\cite{prep_RHC}, as this mode is not sensitive to CKM factors. The observables $H$ and $S$ are well-approximated by
 \begin{equation}
 \label{eq:SHphi}
  \HH_{B_s \to \phi (f_1)\ga} \simeq    2   \{   \pm (\Delta_R \cos ( \phi_R) +  \hat{m}_s  ) - \Rea [\epsQCD{c}_{\phi(f_1),R} ] \} \,,  \quad S_{B_s \to \phi(f_1)  \ga} \simeq    2  \{ \pm \Delta_R \sin ( \phi_R)  \}    \;.
 \end{equation}
The vanishing of $S_{B_s \to \phi(f_1)\ga} \simeq 0$ in the \ac{SM}
comes from the cancelation of all weak phases involved, and this quantity is therefore a null test for weak 
phases of \ac{RHC}.
From Eq.~\eqref{eq:SHphi}, we obtain the remarkable equation 
\begin{equation}
\label{eq:Hcharm1}
  \HH_{\phi\ga} +  \HH_{\fone\ga}  \simeq
- 2 {\textrm{Re}}[ \epsQCD{c}_{\phi,R} +  \epsQCD{c}_{\fone,R}   ]  =
- 2 {\textrm{Re}}[ \epsQCD{c}_{\phi,R}] ( 1 +  \RR_{\fone,\phi}^c    )  \;,
\end{equation}
where the \ac{SD} physics drops out. Its \ac{SD}-sensitive  counterpart is 
\begin{alignat}{2}
\label{eq:HRHC1}
 \Delta_R  \cos(\phi_{\Delta_R}) =  \frac{1}{4} ( 
    \HH_{\phi\ga} -  \HH_{\fone\ga} )  + 
\frac{1}{2} \Rea [ \epsQCD{c}_{\phi,R}  - \epsQCD{c}_{\fone,R}]  
 - \hat{ m}_s
    \;.
\end{alignat}
 In Eq.~\eqref{eq:Hcharm1}, 
$\RR_{A,V}^i \equiv \textrm{Re} [\epsQCD{i}_{A,R}]/\textrm{Re} [ \epsQCD{i}_{V,R}]$ is 
the more refined version of Eq.~\eqref{eq:Rschema}, in that it includes the information on the 
flavour of the four-quark operator from which it derives. The main points are as follows:
\begin{itemize}
\item Eq.~\eqref{eq:Hcharm1} shows that one can measure the sum of the  \ac{LD} contributions without
compromise from \ac{RHC} or \ac{SM} \ac{SD} physics, owing to the previously-mentioned exact form 
factor relation $T_1(0) = T_2(0)$.
\item One can extract the \ac{LD} parts of the individual modes, entering Eq.~\eqref{eq:HRHC1}, by an analytic 
prediction of $\RR_{\fone,\phi}^c$. We again stress that it is not the value (deviation from unity) but the error on 
$\RR_{\fone,\phi}^c$ which is important. By flavour symmetries it is clear that the measurement of a single axial vector meson 
can reveal valuable information on the size of \ac{LD} contributions.
\item Making the last point more concrete, an error of $20\%$ on $\RR_{\fone,\phi}^c$, assuming a perfect measurement, allows us to extract ${\textrm{Re}}[ \epsQCD{c}_{\phi,R}]$ to $10\%$. This is a much-improved 
situation as compared to an \emph{a priori} computation \cite{prep_charm}. 
\item These methods apply straightforwardly to charm physics \cite{deBoer:2018zhz}, and can be extended to $B \to V\ell \bar{\ell}$
at low $q^2$, although this will require taking into account that the exact form-factor relation $T_1(0) = T_2(0)$ no longer holds \cite{BSZ2015}. 
In particular, the real and imaginary parts of the angular moment $\mathbb{G}_2^{2,2}$, equivalent to $P_1 = A_T^{(2)}$ and $P_3$ respectively, also exhibit the required linear dependence on 
the right-handed amplitude \cite{GHZ2016,KM05,Becirevic:2011bp}. Measuring the analogues of $\eps_R$ in this channel allows to cross-check the \ac{LD} theory input into the anomalous angular $B \to K^* \mu^+ \mu^-$ measurement e.g. $P_5'$ \cite{ANSS2017,Ciuchini:2017mik,CCDMV2017,Arbey:2018ics}.  
\end{itemize}

\section{Conclusions}

In this work, we have advocated that \acl{LD} effects contaminating searches for \acl{RHC} in $B \to V \ga  (\ell \bar{\ell})$ decays can be controlled by considering the corresponding parity-doubler decay mode $B \to A \ga  (\ell \bar{\ell})$. In the limit where the chiral symmetry is restored, the  V-A contributions to the right-handed amplitude come with the opposite sign between these two channels. This can be applied phenomenologically by combining observables, as shown explicitly for example in measurements of time-dependent CP asymmetry in $B_s \to \phi (f_1) \ga$ in Eqs.~\eqref{eq:Hcharm1} and \eqref{eq:HRHC1}, to extract and measure ratios of \acl{LD} contributions. In turn this can lead to a cleaner extraction of \acl{NP} contributions to \acl{RHC}.

It is again important to stress that the main benefit is not in the prediction of the \acl{LD} ratio itself, but the reduced theoretical uncertainty that results. This is also useful in resolving the current tension between predictions of the size of \acl{LD} charm loop contaminations in exclusive \cite{KSW1995,BZ06CP} and inclusive $B \to X_s \ga$ decays \cite{GGLZ2004}, where the inclusive contamination was estimated to be roughly an order of magnitude larger.

Corrections to the results in this work, applying beyond the symmetry limit, can still be understood systematically by investigating the symmetry relations between vector and axial mesons, such as the parameters entering their light-cone distribution amplitudes \cite{prep_charm,prep_da}. This allows the approach we advocate above to be applied in real-world experimental searches, with good prospects at Belle II and LHCb \cite{Li:2008qma,Aaij:2014kwa}.

\section*{Acknowledgments}

We are grateful to Tadeusz Janowski and Marco Pappagallo for suggestions on the manuscript.
RZ wishes to thank the Moriond participants for a pleasant atmosphere, good discussions 
and atmosphere. JG is supported by an STFC studentship (grant reference ST/K501980/1).

\section*{References}

\bibliographystyle{utphys}
\bibliography{References_Moriond}

\end{document}